\documentclass[12pt,preprint]{aastex}
\usepackage{amsmath,graphicx,epsfig,xspace}
\usepackage[breaklinks=true,draft=true]{hyperref}
\newcommand{\eg}{{\it e.g.\/}}

\newcommand{\etal}{{\it et al.\/}}
\newcommand{\cf}{{\it c.f.\/}}

\newcommand{\azeus}{\textsf{AZEuS}\xspace}
\newcommand{\zeus}{\textsl{ZEUS-3D}\xspace}
\newcommand{\del}{\ensuremath{\partial}}
\newcommand{\dgr}{\ensuremath{^{\circ}}}
\shorttitle{Jets on observational length scales}
\shortauthors{Ramsey \& Clarke}
\slugcomment{Accepted for publication in ApJL}
\received{}
\bibpunct[; ]{(}{)}{;}{a}{,}{,}
\title{Simulating protostellar jets simultaneously at launching and observational scales}
\author{Jon P.\ Ramsey and David A.\ Clarke\altaffilmark{1}}
\altaffiltext{1}{Institute for Computational Astrophysics, Department of Astronomy \& Physics, Saint Mary's University, Halifax, Nova Scotia, Canada B3H 3C3.}
\begin{abstract}
We present the first 2.5-D MHD simulations of protostellar jets that include both the region in which the jet is launched magnetocentrifugally at scale lengths $<0.1$ AU, and where the propagating jet is observed at scale lengths $>10^3$ AU.  These simulations, performed with the new AMR-MHD code \azeus, reveal interesting relationships between conditions at the disc surface, such as the magnetic field strength, and direct observables such as proper motion, jet rotation, jet radius, and mass flux.  By comparing these quantities with observed values, we present direct numerical evidence that the magnetocentrifugal launching mechanism is capable, by itself, of launching realistic protostellar jets.
\end{abstract}
\begin{document}
\maketitle
\section{Introduction}
\label{sec:intro}
Jets and outflows from protostellar objects are fundamental aspects of the current star formation paradigm, and are observed anywhere star formation is ongoing.  The mechanism proposed by \citet{bp82}, in which jets are launched from accretion discs by gravitational, magnetic, and centrifugal forces, has been extensively studied numerically (\eg, \citealt{us85, megpl97, op97a, op97b, op99, klb99, vjo02, vR03, ocp03, pf10, snopc10}).  By treating the accretion disc as a boundary condition (\eg, \citealt{ukrcl95}), one can study jet dynamics independently of the disc (\eg, \citealt{pofb07_ppv}) though, in order to resolve the launching mechanism, numerical simulations have not followed the jet beyond 100 AU (\eg, \citealt{alkb05}).

In stark contrast, protostellar jets are $\gtrsim10^4$ AU long \citep{brd07_ppv}, and only recently have observations reached within 100 AU of the source (\eg, \citealp{hep04,cbp08}).  This large scale difference between observations and simulations makes direct comparisons difficult and, in this work, we aim to close this gap.  We present axisymmetric (2.5-D) simulations of protostellar jets launched from the inner AU of a Keplerian disc, and follow the jet well into the observational domain (2500 AU).  These calculations allows us to address the efficacy of the magnetocentrifugal mechanism, and to relate conditions near the disc with directly observable properties of the jet.

The simulations presented herein are performed with an adaptive mesh refinement (AMR) version of \zeus \citep{c96,c10} called \azeus (Adaptive Zone Eulerian Scheme).  The \zeus family of codes are among the best tested, documented, and most widely used astrophysical MHD codes available, though this is the first attempt to couple \zeus with AMR\footnote{\textsl{ENZO\/}, a hybrid N-body Eulerian code \citep{osheaetal04}, links AMR with the \emph{hydrodynamical} portion of \textsl{ZEUS-2D}.}.  We have implemented the block-based method of AMR detailed in \citet{bc89} and \citet{bbsw94}.  Significant effort was spent minimising errors caused by passing waves across grid boundaries, which is of particular importance to this work.  A full description of the code and the changes required for AMR on a fully-staggered mesh will appear in Ramsey \& Clarke (in preparation).
\section{Initialisation}
\label{sec:method}
Observationally, the inner radius of a protostellar accretion disc, $r_{\rm i}$, is between 3--5 $R_{\rm *}$ \citep{chs00} and, for a typical T Tauri star ($M=0.5\,M_{\odot}$, $R_*=2.5R_{\odot}$), $r_{\rm i}=0.05$ AU.  Thus, following \cite{op97a}, we initialise a hydrostatic, force-free atmosphere surrounding a $0.5\,M_{\odot}$ protostar coupled to a rotating disc with $r_{\rm i}=0.05$ AU.  However, unlike \citeauthor{op97a} we use an adiabatic equation of state that conserves energy across shocks rather than an isentropic polytropic equation of state, as the distinction becomes important for supermagnetosonic flow \citep{ocp03}.

We solve the equations of ideal MHD\footnote{\azeus solves either the total or internal energy equation.  We chose the latter because positive-definite pressures trump strict conservation of energy in these simulations; see \citet{c10}.} ($\gamma=5/3$) over a total domain of $4096\,{\rm AU}\times256\,{\rm AU}$.  To span the desired length scales, nine nested, static grids (refinement ratio 2) are initialised each with an aspect ratio of 4:1 (16:1 for the coarsest grid only) and bottom left corner at the origin.  Our finest grid has a domain $4\,{\rm AU}\times1\,{\rm AU}$ and a resolution $\Delta z=r_{\rm i}/8=0.00625\,{\rm AU}$ which we find sufficient to resolve the launching mechanism.  Thus, the effective resolution for the entire domain is $>26$ billion zones.  The simulation highlighted in \S \ref{sec:results} was run to $t=100$~yr with an average time step in the finest grid of $\sim3$ \emph{minutes} and thus $\sim18$ million time steps.

During the simulations, a thin region of low velocity and high poloidal magnetic field, $B_{\rm p}=\sqrt{B_z^2+B_r^2}$, develops along the symmetry axis, the edge of which is defined by a large gradient in the toroidal magnetic field, $\partial_rB_{\varphi}$.  Insufficient resolution of $\partial_rB_{\varphi}$ can lead to numerical instabilities, and grids are added dynamically whenever this gradient is resolved by fewer than five zones.
\subsection{The atmosphere}
\label{sub:atmos}
The atmosphere is initialised in hydrostatic equilibrium (HSE; $v_z=v_r=v_{\varphi}=0$).  Because the LHS of the equation governing HSE,
\begin{equation}
  \label{eq:hseeqn}
  \nabla{p}+\rho\nabla\phi=0,
\end{equation}
is not a perfect gradient, differencing it directly on a staggered-mesh can commit sufficient truncation error to render the atmosphere numerically unstable.  Thus, we replace $\nabla\phi$ with the corresponding poloidal gravitational acceleration vector,
\begin{equation}
\label{eq:hseaccel}
  \vec{g}=-\frac{1}{\rho_{\rm h}}\nabla{p_{\rm h}},
\end{equation}
where $\rho_{\rm h}$ and $p_{\rm h}$ are the hydrostatic density and pressure given by:
\begin{equation}
  \label{eq:hserhop}
  \rho_{\rm h}=\rho_{\rm i}\left(\frac{r_{\rm i}}{\sqrt{r^2+z^2}}\right)^{\frac{1}{\gamma-1}}\quad{\rm and}\quad\quad{p}_{\rm h}=\frac{p_{\rm i}}{\gamma}\left(\frac{\rho_{\rm h}}{\rho_{\rm i}}\right)^{\gamma}.
\end{equation}
Here, $\rho_{\rm i}$ and $p_{\rm i}$ are the initial density and pressure at $r_{\rm i}$ and $p\propto\rho^{\gamma}$ is assumed throughout the atmosphere at $t=0$.  In this way, differencing equation (\ref{eq:hseeqn}) maintains HSE to within machine round-off error \emph{indefinitely}.

However, equations (\ref{eq:hserhop}) as given are singular at the origin where truncation errors are significant regardless of resolution.  These errors can launch a supersonic, narrow jet from the origin destroying the integrity of the simulation.  To overcome this problem, we replace the point mass at the origin with a uniform sphere of the same mass and a radius $R_0$, thus modifying the first of equations (\ref{eq:hserhop}) to:
\begin{equation}
\label{eq:hsesmooth}
\left(\frac{\rho_{\rm h}}{\rho_{\rm i}}\right)^{\gamma-1}=\left\{{\begin{array}{ll}{\displaystyle\frac{r_{\rm i}}{\sqrt{r^2+z^2}},}&\mbox{$r^2+z^2\geq{R_0}^2$};\\~&~\\{\displaystyle\frac{r_{\rm i}}{R_0}\,\frac{3{R_0}^2-r^2-z^2}{2{R_0}^2},}&\mbox{$r^2+z^2<{R_0}^2$}.\end{array}}\right.
\end{equation}
If $R_0$ is sufficiently resolved (\eg, four zones), the numerical jet is eliminated.  The resulting ``smoothed potential" is superior to a ``softened potential" since the former has no measurable effects beyond $R_0$. Here, we use $R_0=r_{\rm i}$.

The atmosphere is initialised with the force-free magnetic field used by \citet{op97a}:
\begin{equation}
  \label{eq:aphi}
\begin{array}{rcl}
A_{\varphi}&\!\!\!=\!\!\!&{\displaystyle\frac{B_{\rm i}}{\sqrt{2-\sqrt{2}}}\ \frac{\sqrt{r^2+(z+z_{\rm d})^2}-(z+z_{\rm d})}{r}};\\[18pt]
B_z&\!\!\!=\!\!\!&{\displaystyle\frac{1}{r}\frac{\del\left(rA_{\varphi}\right)}{\del{r}},\quad\quad{B}_r=-\frac{\del{A}_{\varphi}}{\del{z}}},\quad\quad{B}_{\varphi}=0,
\end{array}
\end{equation}
where $A_{\varphi}$ is the vector potential, $z_{\rm d}$ is the disc thickness (set to $r_{\rm i}$), and $B_{\rm i}$ is the magnetic field strength at $r_{\rm i}$, given by: 
\begin{equation}
\label{eq:betai}
  B_{\rm i}=\sqrt{\frac{8\pi{p}_{\rm i}}{\beta_{\rm i}}}.
\end{equation}
Here, $p_{\rm i}$ and $\beta_{\rm i}$ (plasma beta at $r_{\rm i}$) are free parameters.

Finally, to ensure the declining density and magnetic field profiles do not fall below observational limits, we add floor values $\rho_{\rm floor}\sim10^{-6}\rho_{\rm i}$ and $B_{z,{\rm floor}}\sim10^{-5}B_{\rm i}$ (c.f.\ \citealt{bt07}, \citealt{v03}) to equations (\ref{eq:hsesmooth}) and (\ref{eq:aphi}).  By imposing HSE and the adiabatic gas law at $t=0$, a floor value on $\rho$ imposes effective floor values on $\vec{g}$ and $p$ as well.
\subsection{Boundary Conditions}
\label{sub:bcs}
In the accretion disc ($z\leq0$, $r\geq{r}_{\rm i}$), $v_{\varphi}=v_{\rm K}=\sqrt{GM_*/r}$, the Keplerian speed, and $v_z=\zeta v_{\rm K}=10^{-3}v_{\rm K}$ is an ``evaporation speed" at the disc surface.  The disc and atmosphere are initially in pressure balance with a density contrast $\eta=\rho_{\rm disc}/\rho_{\rm atm}=100$, while $\vec{B}$ is initialised using equations (\ref{eq:aphi}).

Following \citet{klb99}, $\rho,~p$, and $v_z$ are held constant, $v_r=v_zB_r/B_z$, $v_{\varphi}=v_{\rm K}+v_zB_{\varphi}/B_z$, $E_z(-z)=E_z(z)$ (where $\vec{E}=\vec{v}\times\vec{B}$ is the induced electric field), $E_r(0)=v_{\rm K}B_z(0)$, $E_r(-z)=E_r(0)-E_r(z)$, $E_{\varphi}(0)=0$, and $E_{\varphi}(-z)=-E_{\varphi}(z)$.  Since $v_z$ is sub-slow, these conditions are formally over-determined and $p$ should probably be allowed to float.  Indeed, we allowed $p$ to be determined self-consistently in test simulations, and found only minor quantitative differences in the jet since the pressure gradient is only about 1\% of the net Lorentz force at the disc surface.  However, allowing $p$ to float in the boundary caused undue high temperatures in the disc, and thus small time steps.  Therefore, the simulation proceeds more rapidly but otherwise virtually unchanged when $p$ is maintained at its initial value.

Inside $r_i$ ($z\leq0$), we apply reflecting, conducting boundary conditions ($\vec{J}=\nabla\times\vec{B}\neq0$).  Thus, $\rho,~p$, and $\vec{v}$ are reflected across $z=0$, and magnetic boundary conditions are set according to $E_z(-z)=-E_z(z)$, $E_r(-z)=E_r(z)$, and $E_{\varphi}(-z)=E_{\varphi}(z)$. At $z=0$, $E_r$ and $E_{\varphi}$ are evolved using the full MHD equations.

Finally, we use reflecting boundary conditions along the $r=0$ symmetry axis, and outflow conditions along the outermost $r$ and $z$ boundaries.
\subsection{Scaling Relations}
\label{sub:scaling}
From equation (\ref{eq:hseeqn}) and the adiabatic gas law, one can show:
\begin{equation}
\label{eq:csvk}
c_{\rm s}^2=\gamma\frac{p}{\rho}=\left(\gamma-1\right)\frac{GM_{\rm *}}{R}=\left(\gamma-1\right)v_{\rm K}^2,
\end{equation}
where $R$ is the spherical polar radius.  From equations (\ref{eq:betai}), (\ref{eq:csvk}), and the ideal gas law ($p=\rho{kT}/\langle{m}\rangle$, where $\langle{m}\rangle$ is half a proton mass), we derive the following scaling relations to convert from unitless to physical quantities:
\begin{eqnarray}
\label{eq:pscale}
&\displaystyle p_{\rm i}=\left(160~{\rm dyne~cm^{-2}}\right)\left(\frac{\beta_{\rm i}}{40}\right)\left(\frac{B_{\rm i}}{10\,{\rm G}}\right)^2;&\\[3pt]
\label{eq:rhoscale}
&\displaystyle \frac{\rho_{\rm i}}{\langle{m}\rangle}=\left(5.4\times10^{12}~{\rm cm}^{-3}\right)\,\left(\frac{\beta_{\rm i}}{40}\right)\,\left(\frac{B_{\rm i}}{10\,{\rm G}}\right)^2\left(\frac{r_{\rm i}}{0.05\,{\rm AU}}\right)\left(\frac{0.5\,M_{\odot}}{M_{\rm *}}\right);&\\[3pt]
\label{eq:tempscale}
&\displaystyle T_{\rm i}=(2.2\times10^5~{\rm K})\,\left(\frac{0.05\,{\rm AU}}{r_{\rm i}}\right)\left(\frac{M_{\rm *}}{0.5\,M_{\odot}}\right);&\\[3pt]
\label{eq:csvkscale}
&\displaystyle c_{\rm s,i}=(77~{\rm km\,s}^{-1})~\left(\frac{0.05\,{\rm AU}}{r_{\rm i}}\right)^{1/2}\left(\frac{M_{\rm *}}{0.5\,M_{\odot}}\right)^{1/2};&\\[3pt]
\label{eq:tscale}
&\displaystyle \tau_{\rm i}=\frac{r_{\rm i}}{c_{\rm s,i}}=\left(9.7\times10^4~{\rm s}\right)~\left(\frac{r_{\rm i}}{0.05\,{\rm AU}}\right)^{3/2}\left(\frac{0.5\,M_{\odot}}{M_{\rm *}}\right)^{1/2},&
\end{eqnarray}
for $\gamma=5/3$.  Note that $\beta_{\rm i}$ is the only free parameter varied in this work.
\section{Results for $\beta_{\rm i}=40$}
\label{sec:results}
Figure \ref{fig:zoom} depicts a jet with $\beta_{\rm i}=40$ at $t\simeq100$~yr from the highest resolution grid near the disc surface (bottom panel) to the coarsest grid in which the jet has reached a length of just under 2500 AU (top panel)\footnote{Time-lapse animations are available at \url{http://www.ica.smu.ca/zeus3d/rc10/}.}.  A few features worth noting include:

\begin{figure}[t]
  \begin{center}
    \includegraphics[width=1.01\textwidth,clip=true]{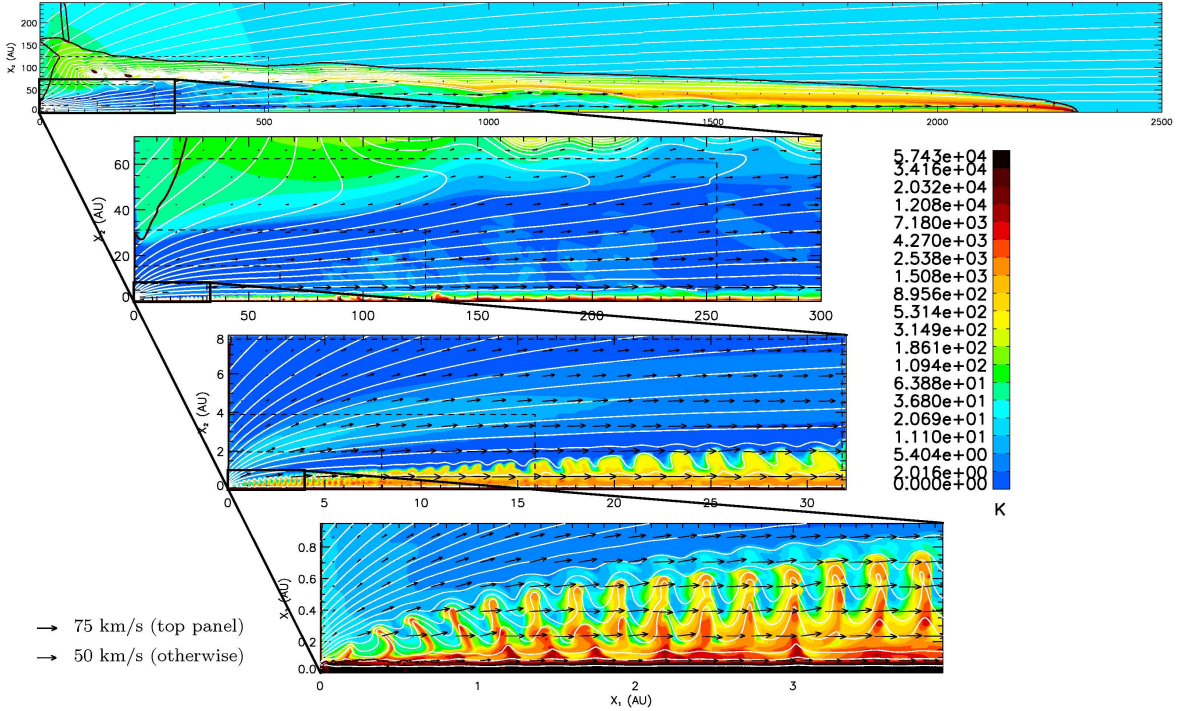}
  \end{center}
  \caption{\label{fig:zoom} Nested images of a $\beta_{\rm i}=40$ jet at $t=100$~yr.  Colours indicate temperature, white contours magnetic field lines, maroon contours the slow surface, and arrows the velocity.  Dashed lines denote grid boundaries.}
\end{figure}

\begin{itemize}
  
  \item When $\theta<60\degr$ (angle between $\vec{B}_{\rm p}$ and disc surface), \cite{bp82} show that cold gas near the disc is launched into a collimated outflow.  Here, $\theta<60\dgr$ for all $r>r_{\rm i}$, but significant outflow is limited to inside the point where the slow surface intersects the disc ($r_{\rm j,d}\sim30$ AU = jet radius at the disc; second panel from top).  Below $r_{\rm j,d}$, cold disc material has moved onto the grid and accelerated into the outflow.  Above $r_{\rm j,d}$, the weak magnetic field has yet to drive enough disc material onto the grid to displace the hot atmosphere, and outflow is stifled.  While $r_{\rm j,d}$ gradually increases with time, the majority of mass flux \emph{originating from the disc} is driven within $r_{\rm i}<r<10\,r_{\rm i}$ (0.5 AU; bottom panel).
  
  \item Jet material becomes super-fast ($M_{\rm f}\lesssim5$) within a few AU of the disc, and the boundary between jet and entrained ambient material is defined by a steep temperature gradient (contact discontinuity; second panel).  Portions of the original atmosphere, which remain virtually stationary throughout the simulation, are still visible above and ahead of the bow shock (top panel).

  \item At large distances from the disc ($\gtrsim500$ AU; top panel), the dynamics of the jet become dominated by $B_{\varphi}$, and the jet is led by an essentially ballistic, magnetic ``nose-cone" with a Mach number of $\sim10$ (\eg, \citealp{cnb86}).  Still, $B_{\varphi}$ is a small fraction ($10^{-3}$) of $B_{\rm i}$, consistent with \cite{hfvb07}.

  \item The knots dominating the bottom panel (\cf, \citealt{op97b}) are produced by the nearly harmonic oscillation of $\vec{B}_{\rm p}$ in $r_{\rm i}<r<2\,r_{\rm i}$, whereby $\theta$ fluctuates between $55\dgr$ and $65\dgr$ with a period $\sim30\,\tau_{\rm i}$.  These oscillations result from the interplay between in-falling material along the symmetry axis, and under/over pressurisation near the central mass.  The knots are denser and hotter than their surroundings, and bound by magnetic field loops.  They occupy a region within $\sim2$ AU of the symmetry axis, and gradually merge to form a continuous and narrow column of hot, magnetised material\footnote{The knots are resolved by 10--20 zones when they merge, and thus their merger is unlikely related to the ever-decreasing resolution of the nested grids.} (third panel).  As such, they are unlikely to be the origin of the much larger-scale knots observed in some jets (\eg, HH111; \citealp{rnrghbc02}).
  
\end{itemize}

Further details of this and other simulations of protostellar jets are left to a future paper, and we focus here on a few properties directly comparable with observations.

\section{Comparing simulations and observations}
\label{sec:compare}
Table \ref{tab:genchar} summarises a few observational characteristics of protostellar jets.  To connect these attributes to conditions in the launching region, we have performed a small parameter survey in $\beta_{\rm i}$, and made numerical measurements of the quantities in Table \ref{tab:genchar}.  Variation of other parameters (such as $\zeta$ and $\rho_{\rm i}$) is left to future work.

\begin{table}[t]
  \begin{center}
  \begin{tabular}{c|c}
    \tableline\tableline
    proper motion (km s$^{-1}$) & 100 -- 200 (500 max.)\\
    rotational velocity (km s$^{-1}$) & (5 -- 25) $\pm$ 5 \\
    FWHM jet width (AU) & 30 -- 80 (at 200 AU) \\
    mass-loss rate ($10^{-6}\,M_{\odot}\,{\rm yr}^{-1}$) & 0.01 -- 1 \\
    \tableline
  \end{tabular}
  \caption{\label{tab:genchar} Selected observational characteristics of protostellar jets.  References: \citet{rb01}, \citet{rdbec07_ppv}, \citet{mo07}. }
  \end{center}
\end{table}

Note that $\beta_{\rm i}$ is the initial value of the plasma beta at $r_{\rm i}$, and not the average $\beta$ in the jet.  Indeed, Fig.\ \ref{fig:pm}a demonstrates that at very early time, $\langle\beta\rangle=8\pi\langle{p}\rangle/\langle{B^2}\rangle\lesssim\beta_{\rm i}/5$, where $B^2=B_{\rm p}^2+B_{\varphi}^2$, and where the average is taken over zones that exceed a certain threshold $v_z$ so that only out-flowing jet material is considered. Thus, the magnetic field within the jet is stronger than $\beta_{\rm i}$ would suggest.  Initially, $\langle\beta\rangle$ is dictated by $B_{\rm p}$, but becomes dominated by $B_{\varphi}$ within $\lesssim10$~yr after launch.  As time progresses, $\langle\beta\rangle$ gradually increases but never rises above unity (at least for $t<100$~yr), even for $\beta_{\rm i}\gg1$.  Still, one might speculate from Fig.\ \ref{fig:pm}a that with sufficient time, $\langle\beta\rangle\rightarrow1$ regardless of $\beta_{\rm i}$.

\subsection{Proper motion}
\label{sub:pm}
\begin{figure}[t]
  \begin{center}
    \hspace*{-2.0ex}\includegraphics[scale=0.59]{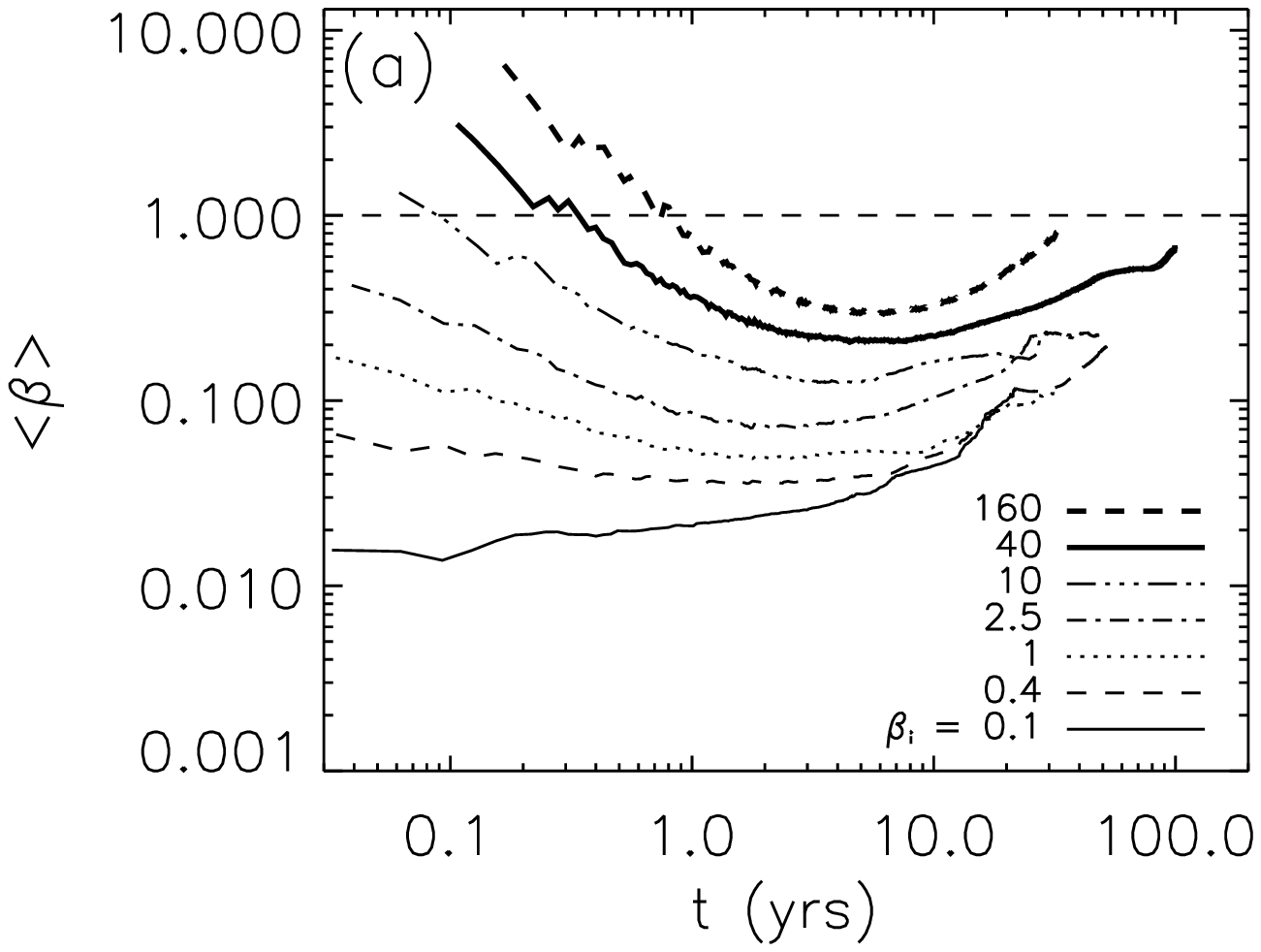}\hfill
    \hspace*{-2.0ex}\includegraphics[scale=0.59]{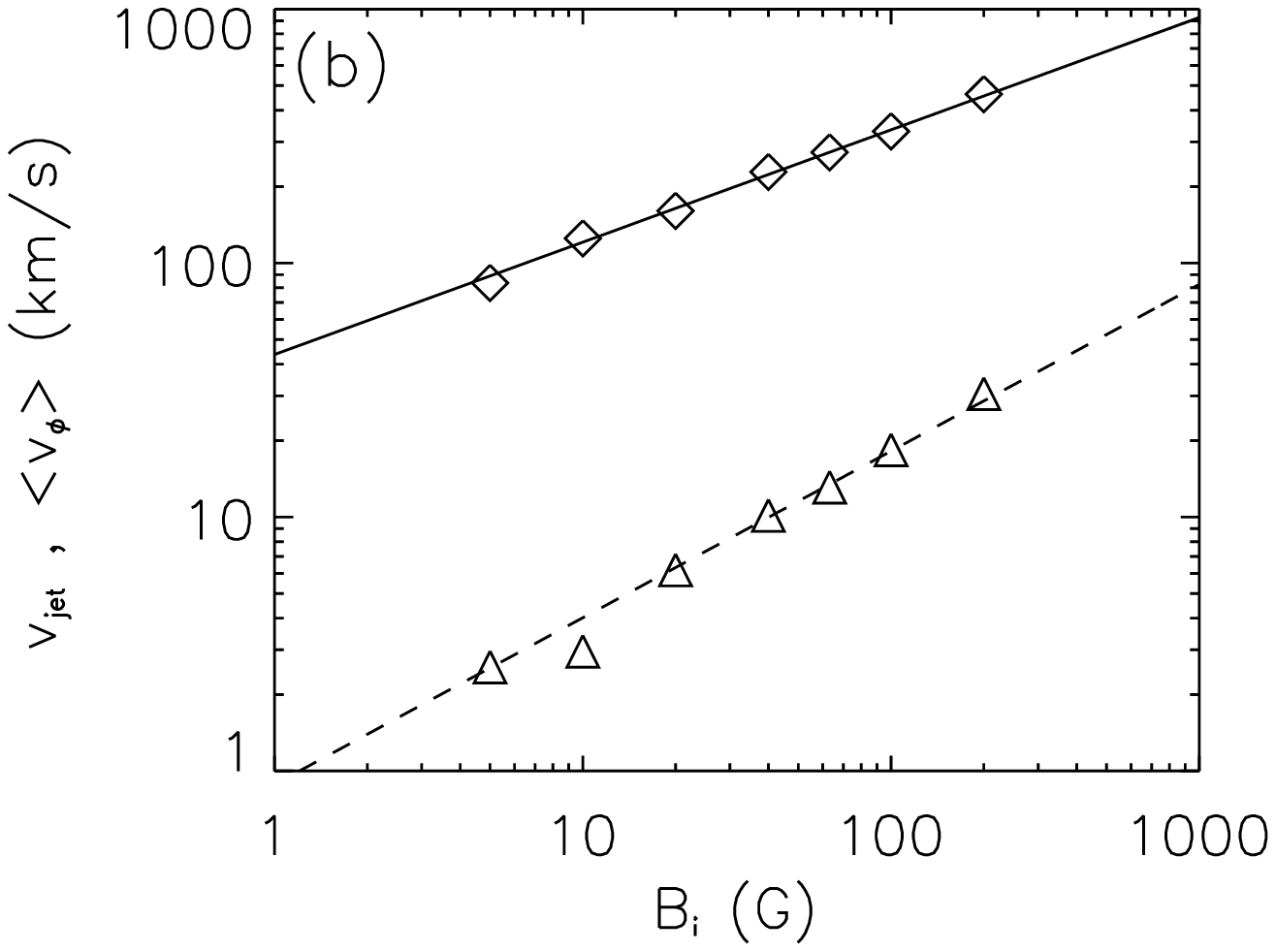}
    \caption{\label{fig:pm} (a) $\langle\beta\rangle$ as a function of time for different $\beta_{\rm i}$.  (b) $v_{\rm jet}$ (diamonds) and $\langle{v}_{\varphi} \rangle$ (triangles) of each jet as a function of $B_{\rm i}$.  Best fit power-law coefficients for these data are $\alpha=0.44\pm0.01$ ($v_{\rm jet}$, solid line) and $0.66\pm0.01$ ($\langle{v}_{\varphi}\rangle$, dashed line).}
  \end{center}
\end{figure}

For $t\gtrsim10$~yr, the velocity of the tip of the jet, $v_{\rm jet}$, is nearly constant and, from Fig.\ \ref{fig:pm}b and Table \ref{tab:simchar}, we find $v_{\rm jet}\propto{B}_{\rm i}^{0.44\pm0.01}$.

To understand this result physically, we begin with the magnetic forces:
\begin{align}
  \label{eq:flforces}
  F_{\parallel} & = -\frac{B_\varphi}{r} \nabla_{\parallel}\left( rB_\varphi \right); \nonumber \\
  F_\varphi & = \frac{B_{\rm p}}{r} \nabla_{\parallel}\left( rB_\varphi \right); \\
  F_{\perp} & = -\frac{B_\varphi}{r} \nabla_{\perp}\left( rB_\varphi \right) + J_\varphi B_{\rm p}, \nonumber
\end{align}
(\eg, \citealt{f97,zfrbm07}) where $\nabla_\parallel,\nabla_\perp$ are the gradients parallel and perpendicular to $\vec{B}_{\rm p}$.  For a given field line, a stronger $B_{\rm p}$ at its ``footprint" in the disc ($r=r_0$) generates a stronger $B_\varphi$ which leads to stronger gradients in $rB_\varphi$ and thus, from equations (\ref{eq:flforces}), greater magnetic forces to accelerate the flow.  In practice, we find that most of the acceleration occurs before the fast point (and not the Alfv\'en point) located at $r=r_{\rm f}$, where $r_{\rm f}$ is a weak function of the field strength at the footprint and thus of $B_{\rm i}$.

Following \citet{s96}, one can show that as a function of the ``fast moment arm" ($\xi\equiv{r}_{\rm f}/r_0$), the poloidal velocity at the fast point is:
\begin{equation}
\label{eq:vffunc}
v_{\rm p,f}=\sqrt{a_{\rm p,f}\,v_{\rm K,0}}\left(\xi^2+\frac{2}{\xi}-3\right)^{1/4}\propto~\sqrt{B_{\rm i}}
\end{equation}
since $a_{\rm p,f}$, the poloidal Alfv\' en speed at the fast point, is roughly proportional to $B_{\rm i}$.  $v_{\rm K,0}=\sqrt{GM_{\rm *}/r_0}$ is the Keplerian speed at the footprint of the field line.  We note that measured values of $v_{\rm p,f}$ in our simulations vary as $B_{\rm i}^{0.5}$ and agree with equation (\ref{eq:vffunc}) to within 1\% so long as the fluid is in approximate steady-state\footnote{Indeed, all four steady-state functions from \citet{s96} remain constant in our simulations to within $\lesssim$ 5\% along steady-state field lines, which we take as validation of our numerical methods.}.

After the poloidal force given by equations (\ref{eq:flforces}) decreases to 1\% of its maximum value ($\gtrsim$ a few $r_{\rm f}$), $v_{\rm p}$ still follows a power law in $B_{\rm i}$ with index $0.52\pm0.04$ and essentially unchanged from equation (\ref{eq:vffunc}).  Nearer the head of the jet where steady state is no longer valid, we find $\langle{v}_{\rm p}\rangle\propto{B}_{\rm i}^{0.45\pm0.02}$ (where the momentum-weighted average is taken across the jet radius), only slightly shallower than equation (\ref{eq:vffunc}).  Thus, while the conditions in the jet have changed, some memory of the steady-state conditions at $r_{\rm f}$ persists.

Finally, $v_{\rm jet}$ (Fig.\ \ref{fig:pm}b and Table \ref{tab:simchar}) is within $\sim 10\%$ of $\langle{v}_{\rm p}\rangle$ near the bow shock and maintains the same power-law dependence on $B_{\rm i}$.  Thus, these jets are essentially ballistic, where the observed jet speed $v_{\rm jet}\propto{B}_{\rm i}^{0.44\pm0.01}$.  In short, all measures of jet speed increase with $B_{\rm i}$, a trend that agrees with \citet{alkb05} who find for much less evolved jets, $v_{\rm p} \propto B_{\rm i}^{1/3}$.

\begin{table}[t]
  \begin{center}
  \begin{tabular}{c|rrrrrrr||c}
    $\beta_{\rm i}$ & 160 & 40 & 10 & 2.5 & 1.0  & 0.4 & 0.1 &  \\
    $B_{\rm i}$ (G) & 5   & 10 & 20 & 40  & 63.2 & 100 & 200 & $\alpha$ \\
    \tableline\tableline
    $v_{\rm jet}$ (km s$^{-1}$) & 84 & 125 & 161 & 230 & 270 & 330 & 460 & $0.44\pm0.01$ \\
    $\langle v_\varphi\rangle$ (km s$^{-1}$) & 2.6 & 3.0 & 6.2 & 10.1 & 13.1 & 18.4 & 31 & $0.66\pm0.01$ \\
    $2 \, r_{\rm jet}$ (AU) & 21 & 40 & 60 & 85 & 94 & 104 & 130 & $0.35\pm0.04$ \\
    $\dot{M}_{\rm jet}$ ($10^{-6}M_\odot~{\rm yr}^{-1}$) & 0.44 & 1.9 & 2.8 & 4.2 & 6.9 & 10.1 & 17.9 & $0.92\pm0.09$ \\
    \tableline
  \end{tabular}
  \caption{\label{tab:simchar} Simulation ``observables" $v_{\rm jet}$ and $\langle{v}_{\varphi}\rangle$ are asymptotic values while $r_{\rm jet}$ and $\dot{M}_{\rm jet}$ are measured at $z=200$~AU and $t=20$~yr.  Uncertainties in $\alpha$ are from the fitting procedure.}
  \end{center}
\end{table}
\subsection{Toroidal velocity}
\label{sub:vphi}
Figure \ref{fig:pm}b and Table \ref{tab:simchar} show $v_\varphi$ averaged over time and the jet volume for $z\geq$100 AU as a function of $B_{\rm i}$.  Like $v_{\rm jet}$, $v_{\varphi}$ asymptotes to a constant value.  The region inside 100 AU is ignored because the torsion Alfv\' en wave at low $z$ has a non-negligible $v_\varphi$, is not part of the jet, and skews our results.  By fitting a power law to these data, we find $\langle{v}_{\varphi}\rangle\propto{B}_{\rm i}^{0.66\pm0.01}$.

Unlike $v_{\rm jet}$, we have not uncovered a rationale for this power law, yet it seems plausible one must exist given the tightness of fit.  Eliminating $B_{\rm i}$ from the power laws for $\langle{v}_{\varphi}\rangle$ and $v_{\rm jet}$, we find that $\langle{v}_{\varphi}\rangle\propto{v}_{\rm jet}^{1.50\pm0.06}$.  To render this a useful observational tool, further work is needed to quantify the effects of other initial conditions such as $\zeta$ and $\rho_{\rm i}$ on both the power law index and the proportionality constant, as well as the effect our simplified disc model may have on conditions in the jet at observational length scales.
\subsection{Jet radius and mass flux}
\label{sub:jetradius}
The jet radius, $r_{\rm jet}$, is defined by the contact discontinuity (steep temperature gradient in the second panel of Fig.\ \ref{fig:zoom}) between shocked jet and shocked ambient material, which in turn is determined by where the radial jet ram pressure balances all external forces.  Since ram pressure increases with $v_{\rm p}$, $r_{\rm jet}$ should increase with $B_{\rm i}$, just as observed in Table \ref{tab:simchar}.  At any given time, we find that $r_{\rm jet}$ varies with $B_{\rm i}$ as a reasonable power law though, unlike $v_{\rm jet}$ or $\langle{v}_{\varphi}\rangle$, the power index is not constant and decreases slowly in time, while $r_{\rm jet}$ itself increases in time, though at an ever-slowing rate. 

The mass flux transported by the jet, $\dot{M}_{\rm jet}$, consists of material from both the disc and the atmosphere.  Unlike previous simulations where jets are typically evolved long after the leading bow shock has left the grid, no part of any bow shock in our simulations reaches the boundary of the coarsest grid.  Thus, each jet continues to entrain material from the atmosphere throughout the simulation at a rate that has a strong dependence on $B_{\rm i}$, as seen in Table \ref{tab:simchar}.  Indeed we find that $\dot{M}_{\rm jet}$ varies with $B_{\rm i}$ as a reasonable power law, with the power index decreasing slowly in time.  As the atmosphere is depleted, the mass flux contribution from the disc (which, by design, is independent of $B_{\rm i}$) becomes more important and the dependence of $\dot{M}_{\rm jet}$ on $B_{\rm i}$ diminishes.
\section{Discussion}
\label{sec:discuss}
We have presented the first MHD simulations of protostellar jets that start from a well-resolved launching region ($\Delta{z}_{\rm min}=0.00625$~AU) and continue well into the observational domain (2500 AU).  On the AU scale, each jet shows the characteristic and near steady-state knotty behaviour first reported by \citet{op97b}, though the origin of our knots is quite different.  On the 1000 AU scale, each jet develops into a ballistic, supersonic ($8\lesssim{M}\lesssim11$) outflow led by a magnetically confined ``nose-cone" \citep{cnb86} and a narrow bow shock, consistent with what is normally observed.

On comparing Tables \ref{tab:genchar} and \ref{tab:simchar}, our simulations comfortably contain virtually all observed protostellar jets on these four important quantities.  We note that these tables would \emph{not} have been in agreement had we stopped the jet at, say, 100 AU and measured these values then.  \emph{It is only because our jets have evolved over five orders of magnitude in length scale that we can state with some confidence that the magnetocentrifugal launching mechanism is, by itself, capable of producing jets with the observed proper motion, rotational velocity, radius, and mass outflow rate.}  Indeed, our jets are still very young, having evolved to only 100 yr, and allowing them to evolve over an additional one or two orders of magnitude in time may still be useful.  For example, it would be interesting to know whether $\langle\beta\rangle$ rises above unity for any of the jets (Fig.\ \ref{fig:pm}a), and thus enter into a hydrodynamically dominated regime.  It would also be interesting to see how long it takes for the power laws in jet radius and mass flux as a function of $B_{\rm i}$ to reach their asymptotic limits.

Our jet widths tend to be higher than those observed, particularly when one considers that the values for $r_{\rm jet}$ in Table \ref{tab:simchar} are at $t=20$~yr\footnote{Some simulations had not reached $t=100$~yr at the time of this writing.}, and that $r_{\rm jet}$ continues to grow in time (\eg, for the $\beta_{\rm i}=40$ jet, $2\,r_{\rm jet}\sim100$~AU by $t=100$~yr).  As our jet radii mark the locations of the contact discontinuity while observed radii mark hot, emitting regions, our widths should be considered upper limits.  That our values \emph{contain} all observed jet widths is a success of these simulations.

Similarly, our numerical mass fluxes are higher than observed values by at least an order of magnitude.  Since observed mass-loss rates account only for emitting material (\eg, in forbidden lines; \citealt{hmr94}), and thus temperatures in excess of $10^4$ K (\citealt[p.\ 104]{dw97_book}), our mass fluxes are necessarily upper limits as well.  Indeed, if we measure our mass fluxes near the jet tip (instead of at 200 AU for Table \ref{tab:simchar}) and restrict the integration to fluid above $10^4$ K, our mass fluxes drop by a factor of 10--100, in much better agreement with Table \ref{tab:genchar}.
\acknowledgements
We thank the referee for timely and helpful comments on the manuscript, Marsha Berger for her AMR subroutines, and Sasha Men'shchikov for early work on \azeus.  Use of MPFIT by C.~B.~Markwardt and JETGET by J.~Staff, M.~A.~S.~G.~J{\o}rgenson, and R.~Ouyed is acknowledged.  This work is supported by NSERC.  Computing resources were provided by ACEnet which is funded by CFI, ACOA, and the provinces of Nova Scotia, Newfoundland \& Labrador, and New Brunswick.

\newpage
\begin{thebibliography}{}
  \bibitem [Anderson \etal(2005)]{alkb05} Anderson, J.\ M., Li, Z.-Y., Krasnopolsky, R., Blandford, R.\ D., 2005, ApJ, 630, 945.
  \bibitem [Bally, Reipurth, \& Davis(2007)]{brd07_ppv} Bally, J., Reipurth, B., Davis, C.\ J., 2007, in Protostars and Planets V, eds.\ B.\ Reipurth, D.\ Jewitt, K.\ Keil (Tucson: Univ.\ of Arizona Press), 215.
  \bibitem [Bell \etal(1994)] {bbsw94} Bell, J., Berger, M., Saltzman, J., Welcome, M., 1994, SIAM J.\ Sci.\ Comput., 15, 127.
  \bibitem [Berger \& Colella(1989)] {bc89} Berger, M.\ J., Colella, P., 1989, JCoPh, 82, 64.
  \bibitem [Bergin \& Tafalla(2007)] {bt07} Bergin, E.A., Tafalla, M., 2007, ARA\&A, 45, 339.
  \bibitem [Blandford \& Payne(1982)BP82]{bp82} Blandford, R.\ D., Payne, D.\ G., 1982, MNRAS, 199, 883.
  \bibitem [Calvet \etal(2000)] {chs00} Calvet, N., Hartmann, L., Strom, S.\ E., 2000, in Protostars and Planets IV, eds. V.\ Mannings, A.\ P.\ Boss, S.\ S.\ Russell (Tucson: Univ.\ of Arizona Press), 377.
  \bibitem [Clarke, Norman, \& Burns(1986)] {cnb86} Clarke, D.\ A., Norman, M.\ L., Burns, J.\ O., 1986, ApJ, 311, L63.
  \bibitem [Clarke(1996)] {c96} Clarke, D.\ A., 1996, ApJ, 457, 291.
  \bibitem [Clarke(2010)] {c10} Clarke, D.\ A., 2010, ApJS, 187, 119.
  \bibitem [Coffey \etal(2008)] {cbp08} Coffey, D., Bacciotti, F., Podio, L., 2008, ApJ, 689, 1112.
  \bibitem [Dyson \& Williams(1997)] {dw97_book} Dyson, J.\ E., Williams, D.\ A., 1997, The Physics of the interstellar medium (2nd ed.; Bristol: IOP Publishing).
  \bibitem [Ferreira(1997)] {f97} Ferreira, J., 1997, A\&A, 319, 340.
  \bibitem [Hartigan, Morse, \& Raymond(1994)] {hmr94} Hartigan, P., Morse, J.\ A., Raymond, J., 1994, ApJ, 436, 125.
  \bibitem [Hartigan, Edwards, \& Pierson(2004)] {hep04} Hartigan, P., Edwards, S., Pierson, R., 2004, ApJ, 609, 261.
  \bibitem [Hartigan \etal(2007)] {hfvb07} Hartigan, P., Frank, A., Varni\' ere, P., Blackman, E.\ G., 2007, ApJ, 661, 910.
  \bibitem [Krasnopolsky \etal(1999)] {klb99} Krasnopolsky, R., Li, Z.-Y., Blandford, R., 1999, ApJ, 526, 631.
  \bibitem [McKee \& Ostriker(2007)] {mo07} McKee, C.\ F., Ostriker, E.\ C., 2007, ARA\&A, 45, 565.
  \bibitem [Meier \etal(1997)] {megpl97} Meier, D.\ L., Edgington, S., Godon, P., Payne, D.\ G., Lind, K.\ R., 1997, Nature, 388, 350.
  \bibitem [O'Shea \etal(2004)] {osheaetal04} O'Shea, B.\ W., Bryan, G., Bordner, J., Norman, M.\ L., Abel, T., Harkness, R., Kritsuk, A., 2004, eprint (arXiv: astro-ph/0403044).
  \bibitem [Ouyed, Clarke, \& Pudritz(2003)] {ocp03} Ouyed, R., Clarke, D.\ A., Pudritz, R.\ E., 2003, ApJ, 582, 292.
  \bibitem [Ouyed \& Pudritz(1997a)]{op97a} Ouyed, R., Pudritz, R.\ E., 1997a, ApJ, 482, 712.
  \bibitem [Ouyed \& Pudritz(1997b)]{op97b} Ouyed, R., Pudritz, R.\ E., 1997b, ApJ, 484, 794.
  \bibitem [Ouyed \& Pudritz(1999)] {op99} Ouyed, R., Pudritz, R.\ E., 1999, MNRAS, 309, 233.
  \bibitem [Porth \& Fendt(2010)] {pf10} Porth, O., Fendt. C., 2010, ApJ, 709, 1100.
  \bibitem [Pudritz \etal(2007)] {pofb07_ppv} Pudritz, R.\ E., Ouyed, R., Fendt, C., Brandenburg, A., 2007, in Protostars and Planets V,  eds.\ B.\ Reipurth, D.\ Jewitt, K.\ Keil (Tucson: Univ.\ of Arizona Press), 277.
  \bibitem [Raga \etal(2002)] {rnrghbc02} Raga, A.\ C., \etal, 2002, ApJ, 565, L29.
  \bibitem [Ray \etal(2007)] {rdbec07_ppv} Ray, T., Dougados, C., Bacciotti, F., Eisl\"offel, J., Chrysostomou, A., 2007, in Protostars and Planets V, eds.\ B.\ Reipurth, D.\ Jewitt, K.\ Keil (Tucson: Univ.\ of Arizona Press), 231.
  \bibitem [Reipurth \& Bally(2001)] {rb01} Reipurth, B., Bally, J., 2001, ARA\&A, 39, 403.
  \bibitem [Spruit(1996)] {s96} Spruit, H.C., 1996, in Evolutionary processes in binary starts, eds.\ R.\ A.\ M.\ J.\ Wijers, M.\ B.\ Davies, C.\ A. Tout, (Dordrecht: Kluwer academic publishers), 249.
  \bibitem [Staff \etal(2010)] {snopc10} Staff, J.\ E., Niebergal, B.\ P., Ouyed, R., Pudritz, R.\ E., Cai, K., 2010, ApJ, 722, 1325.
  \bibitem [Uchida \& Shibata(1985)]{us85} Uchida, Y., Shibata, K., 1985, PASJ, 37, 515.
  \bibitem [Ustyugova \etal(1995)] {ukrcl95} Ustyugova, G.\ V., Kolboda, A.\ V., Romanova, M.\ M., Chechetkin, V.\ M., Lovelace, R.\ V.\ E., 1995, ApJ, 439, 3.
  \bibitem [Vall\' ee(2003)] {v03} Vall\' ee, J.P., 2007, NewAR, 47, 85.
  \bibitem [Vitorino \etal(2002)] {vjo02} Vitorino, B.\ F., Jatenco-Pereira, V., Opher, R., 2002, A\&A, 384, 329.
  \bibitem [von Rekowski \etal(2003)] {vR03} von Rekowski, B., Brandenburg, A., Dobler, W., Shukurov, A., 2003, A\&A, 398, 825.
  \bibitem [Zanni \etal(2007)] {zfrbm07} Zanni, C., Ferrari, A., Rosner, R., Bodo, G., Massaglia, S., 2007, A\&A, 469, 811.
%
\end{thebibliography}
\end{document}